\documentclass[aps,prl,twocolumn, unsortedaddress,superscriptaddress]{revtex4}

\usepackage{graphicx}

\newcommand{\CO}{CO }
\newcommand{\COn}{CO}

\begin{document}

\title{Deceleration of neutral molecules in macroscopic traveling traps}
\author{Andreas Osterwalder}
\email[]{andreas.osterwalder@epfl.ch}
\affiliation{Fritz-Haber-Institut der Max-Planck-Gesellschaft, Faradayweg 4-6, 14195 Berlin, Germany}
\affiliation{Present Address: Ecole Polytechnique F\'ed\'erale de Lausanne, Institut des Sciences et Ing\'enerie Chimique, 1015 Lausanne, Switzerland}
\author{Samuel A. Meek}
\author{Georg Hammer}
\author{Henrik Haak}
\author{Gerard Meijer}
\email[]{meijer@fhi-berlin.mpg.de}
\affiliation{Fritz-Haber-Institut der Max-Planck-Gesellschaft, Faradayweg 4-6, 14195 Berlin, Germany}

\date{\today}

\begin{abstract}
A new type of decelerator is presented where polar neutral molecules are guided and decelerated using the principle of traveling electric potential wells, such that molecules are confined in stable three-dimensional traps throughout.
This decelerator is superior to the best currently operational one \cite{Scharfenberg:2009p3004}, providing a substantially larger acceptance even at higher accelerations.
The mode of operation is described and experimentally demonstrated by guiding and decelerating \CO molecules.
\end{abstract}


\maketitle

Increasing interest in cold and slow neutral molecules has been stimulated by their use in scattering studies, quantum information, spectroscopy, and in particular also by the success of similar studies in atomic physics. 
Some of the most powerful techniques to produce ultracold atoms, most notably laser-cooling, could not yet be applied to molecules, although an interesting step forward has recently been reported \cite{shuman:09}.
Over the past decade, several alternative methods for the production of cold and ultracold neutrals were developed (see \cite{Krems:2009p3148} and references therein). 
One approach is to start from ultracold atoms and assemble them to molecules by laser fields or Feshbach resonances. 
Another one is to start from a cold but fast sample of molecules prepared by a supersonic expansion. 
These molecules, or a portion thereof, are then brought to low laboratory frame-of-reference velocities \cite{Meerakker:2008p4104} via collisions \cite{Elioff:2003p586,Egorov:2002p4143}, or via their interaction with inhomogeneous electric \cite{Bethlem:1999p3059} or magnetic \cite{Vanhaecke:2007p963} fields. 
The latter two methods make use of the Stark (Zeeman) effect which leads to an energy level shift in an electric (magnetic) field. 
In an inhomogeneous field, molecules in states that are shifted to higher (lower) energies in increasing field feel a force to regions with lower (higher) field. 
A field that increases in the propagation direction of the molecules can therefore be used to decelerate low-field-seeking molecules.

Stark deceleration of polar molecules has emerged as one of the particularly powerful methods for the preparation of cold neutrals \cite{Meerakker:2008p4104}. 
It has been applied, e.g., to the investigation of collisions between Xe atoms and a beam of OH radicals with well-defined low and tunable velocities \cite{Gilijamse:2006p2988}, to high-resolution spectroscopy of OH \cite{hudson:06}, and for an accurate determination of excited-state lifetimes \cite{vandeMeerakker:2005p3868}.

Stark decelerators are generally operated by rapidly switching electric fields between two static configurations. This generates a traveling effective potential well that moves along with the molecules in the beam and can be used to reduce the kinetic energy of the molecules in a quasi-continuous fashion \cite{Bethlem:2000p3060}.
In a Stark decelerator the same electrodes and electric fields that are used for deceleration are also used to periodically refocus the molecules.
As a result, underfocusing of molecules that oscillate closely around the velocity of the traveling well, as well as overfocusing at low absolute velocities are difficult to avoid and lead to losses of molecules from the decelerator \cite{vandeMeerakker:2005p3869,Sawyer:2008p4144}.
Very low velocities are of particular interest in applications where the molecules are loaded into a stationary trap, which thus far has been accompanied with large losses.
In addition, the coupling between the transverse and longitudinal motion leads to losses.
It has been demonstrated that this can be cured by using a different mode of operation \cite{Scharfenberg:2009p3004,vandeMeerakker:2005p3869}, namely using additional transverse focusing stages. 
Unfortunately, this not only increases the length of the decelerator by a factor of three, but it also raises the minimum forward velocity below which the molecules are overfocused by the same factor. 
Nevertheless, this operation mode provides the largest decelerator acceptance to date \cite{Scharfenberg:2009p3004}.
The acceptance of a decelerator \emph{i.e.}, the volume in six-dimensional momentum-position phase space that can be controllably transmitted through the device, is the most quantitative way to describe its performance and is used here throughout.

In the present paper a new type of Stark decelerator is presented where a genuine moveable three-dimensional trap is implemented. 
This trap, which has cylindrical symmetry, is loaded by initially moving it at the speed of the molecules in the beam. 
Bringing the molecules into stationary traps will then simply be to bring the originally moving traps to standstill, and the problems at low velocities mentioned above will be completely absent. 
Despite being only a first implementation in which deceleration to standstill is not yet possible, the present decelerator already has an acceptance that is superior to that of any currently existing one. 

The method draws upon the basic principle implemented previously for a chip decelerator \cite{Meek:2009p3015}: a periodic array of electrodes is used to produce a spatially modulated electrostatic potential that leads to the desired field minima.
The correct temporal modulation of the voltages leads to a continuous motion of the minima along the structure.
In contrast to the two-dimensional micro-structured decelerator with tubular traps \cite{Meek:2009p3015}, the present decelerator has cylindrical symmetry, and it has similar dimensions as existing Stark decelerators.

\begin{figure}
 \includegraphics[width=3in]{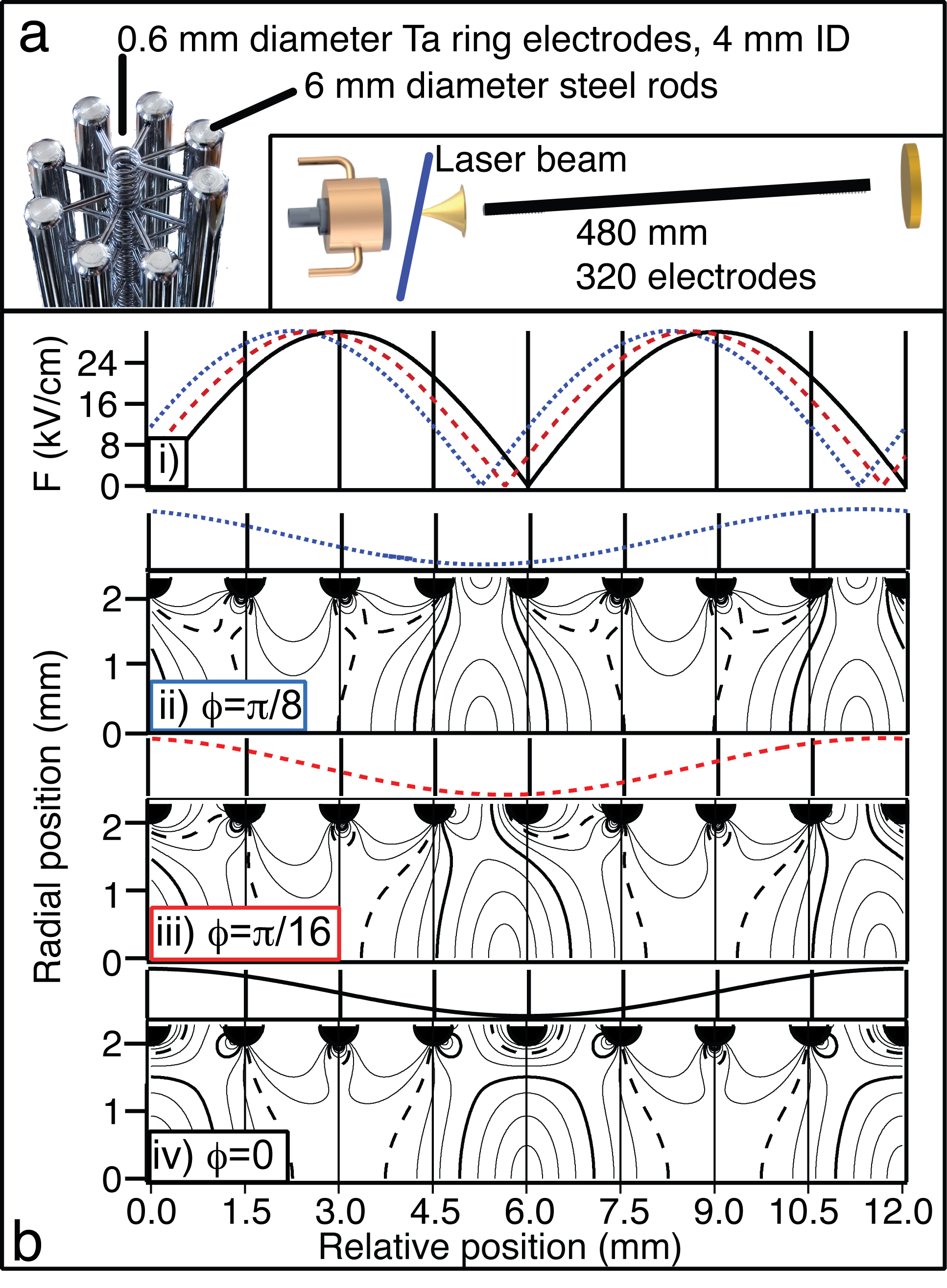}%
\caption{\label{expt}a) Experimental setup and photograph of the last $\sim$40 mm of the decelerator structure. b) i): electric field magnitude along the decelerator axis; ii)-iv): contour plots of the electric field magnitude for three values of $\phi$, and form of the electrostatic potential along the decelerator axis.}
 \end{figure}
The principle of operation is best understood by picturing the electric field formed inside a tube with a cosine-potential applied to the walls:
Longitudinally, that potential follows the same form on the symmetry axis as on the walls while radially, it is described by a zero order modified Bessel function of the first kind.
The longitudinal electric field is zero at the extrema of the cosine, and the radial component of the field is always zero on the symmetry axis.
Around these points, electric field traps are formed with saddle points in between.
For the experimental implementation, the tube is replaced by ring electrodes that sample the tube walls. 
The rings are connected periodically, and the applied voltages sample a cosine. 
Fig.\ \ref{expt}b illustrates the principle of operation for an electrode periodicity of 8 and a period length of 12 mm, as used in the present study.
The contour plots show cross-sections through the cylindrical potential.
The cylinder axis is at the bottom of each panel, and the electrodes are the half-circles at the top.
Contours show the electric field magnitude when a cosine with 16 kV peak-to-peak (p-p) amplitude is applied as shown in small panels above each of the contour plots.
In panel iv, the potential is such that the minimum of the cosine is on the electrode at 6 mm.
In panels iii and ii, the minimum is offset by -0.375 mm and -0.75 mm, respectively.
In each panel, the dashed bold (solid bold) curve shows the contour line for 28 kV/cm (16 kV/cm), and the contours are spaced by 4 kV/cm.
The one-dimensional form of the field magnitude along the symmetry axis is plotted in panel i.
The voltages on the individual electrodes are given by $V_n=V_0\cos(\phi+2\pi \frac{n}{8})$. 
$V_0$ is the amplitude of the waveform, $n=$1, ..., 8 is the number of the electrode, and $\phi$ an overall phase offset. 
A change of $\phi$ shifts the cosine wave and thus also shifts the traps along the cylinder axis.
A continuous motion of the potential along the structure is obtained by a continuous linear increase of $\phi$ in time.
The required voltages for the individual electrodes become $V_n(t)=V_0\cos(\phi\mathrm{(t)}+n\frac{\pi}{4})=V_0\cos(\int_0^t\omega\mathrm{(\tau)d\tau}+n\frac{\pi}{4})$.
The modulation frequency $\omega$ is directly linked to the trap velocity as v=$\frac{\omega}{2\pi}$L, where L is the period length. 
A constant $\omega$ can be used to guide molecules at a selected velocity through the decelerator.
Chirping $\omega$ leads to acceleration or deceleration of the traps.

The experimental set-up, contained in two differentially pumped high-vacuum chambers, is schematically shown in the inset of fig.\ \ref{expt}a. 
A cooled (140 K) solenoid valve (General Valve Series 99, operated at 10 Hz) generates a supersonic expansion of CO from a 20 \% mix in Kr at 2 bar stagnation pressure. 
A Fourier-limited pulsed ($\sim$5 ns) dye laser crosses the expansion directly before a 1 mm skimmer and excites CO molecules to the metastable $a ^3\Pi_1(v=0,J=1,M\Omega=-1)$ state in an $\sim$1 mm long packet with a central velocity around 300 m/s. 
Directly behind the skimmer, the packet enters the decelerator, which is formed from 320 ring electrodes with a center-to-center distance of 1.5 mm, resulting in a 480 mm long decelerator.
The electrodes have an inner diameter of 4 mm, and are made from polished 0.6 mm diameter Ta wire. 
For practicality, and without noticeable effect on the cylindrical symmetry, the shape of the electrodes resembles tennis rackets that are each attached via the handles to one of 8 polished 6 mm diameter stainless-steel rods. 
The left side of fig.\ \ref{expt}a is a photograph showing the end of the decelerator.
The ring electrodes are in the center while the mounting rods are arranged around them.
These rods are arranged on a circle and mounted using macor insulators.
The individual ring electrodes are aligned on each rod spaced by 12 mm. 
The deceleration potentials are generated by applying eight cosine-waveforms to the eight mounting rods. 
All waveforms are at the same frequency but phase-shifted by $\frac{\pi}{4}$ each. 
The 12 mm periodicity of the electrode structure requires a 25 kHz cosine to guide molecules at 300 m/s. 
The individual high-voltage (HV) waveforms are generated by amplifying the output of an eight-channel arbitrary waveform generator (Wuntronic, DA8150; max. 2 V p-p amplitude) in two steps: For each channel, a high-power amplifier (750 W, Servowatt DCP 780/60) increases the amplitude to 100 V p-p and drives a custom-made HV-transformer to produce the (up to) 20 kV p-p waveform. 
The bandwidth of these transformers is currently limited to $\sim$30 kHz --10 kHz. 
The metastable CO molecules possess an internal energy of $\sim$6 eV and are detected directly on a normal Chevron-configured micro-channel-plate detector, and an entire time-of-flight (TOF) trace can be recorded for every experimental cycle.

\begin{figure}
 \includegraphics[width=3.5in]{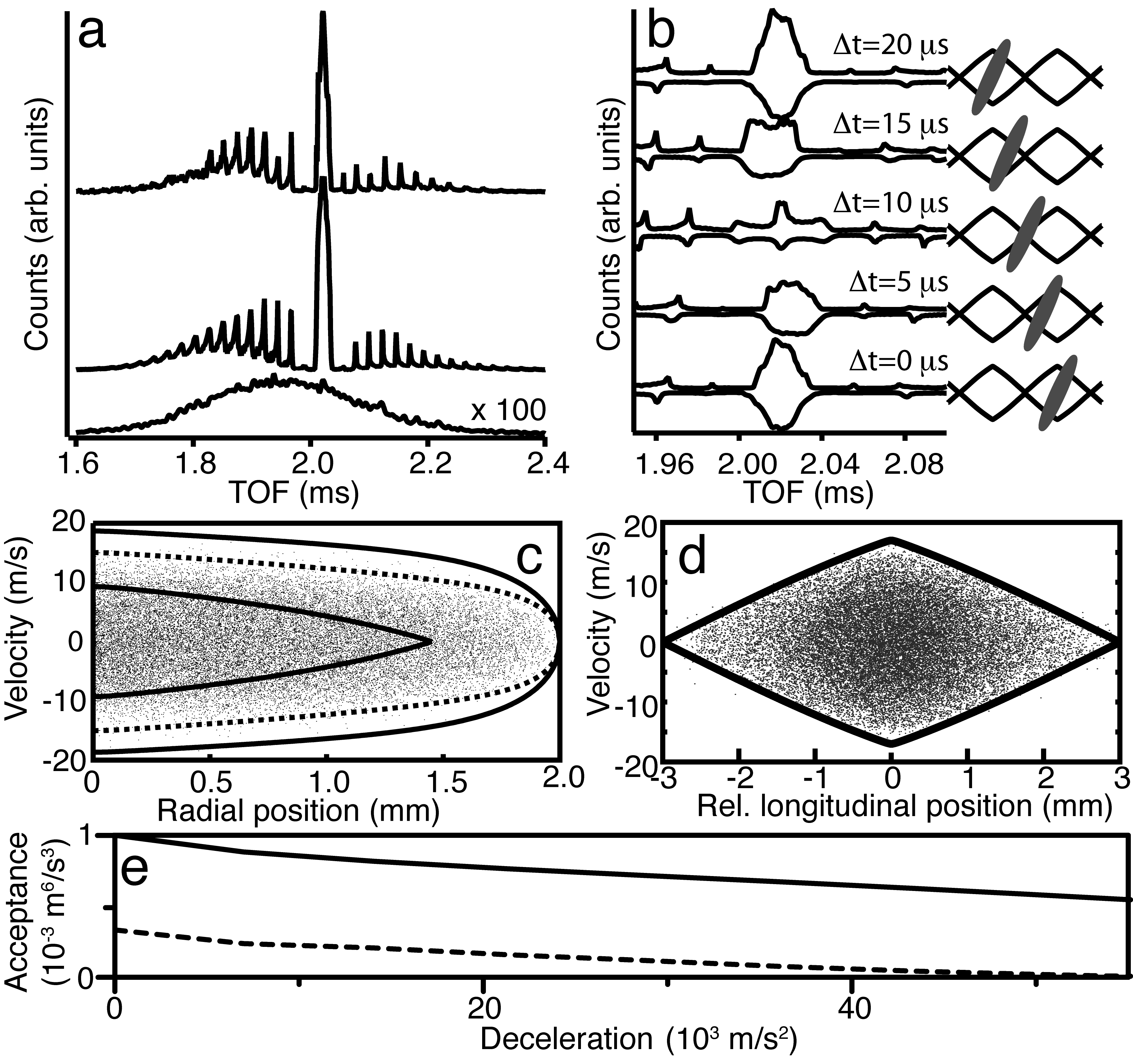}%
\caption{\label{guiding}a) TOF traces for CO at the end of the decelerator without electric fields (bottom trace, scaled up by a factor of 100), and simulated (middle trace) and experimental data (top trace) for guiding at a constant frequency of 24 kHz. b) Central peak from panel a using different values of $\phi$ at pickup, as sketched on the right side of the panel. c) Transverse phase space acceptance. The solid lines represent the separatrices for the two cases $\phi=0$ and $\phi=\pi/8$ (see fig.\ \ref{expt}b), the dashed line is an average of the two extreme cases (see text). d) Longitudinal phase space acceptance. e) Acceptance for OH as a function of deceleration. The solid and dashed lines are the calculated data for the present decelerator and the Stark decelerator described in ref.\ \cite{Scharfenberg:2009p3004}, respectively.}
 \end{figure}
Fig.\ \ref{guiding}a shows a measurement where a 16 kV p-p waveform with a constant frequency of 24 kHz was applied to guide molecules with a velocity of 288 m/s. 
The top trace shows the experimental data, the middle trace a calculated TOF spectrum based on trajectory-simulations.
The original velocity distribution, recorded at the end of the decelerator but without applying any voltages is shown at the bottom of fig.\ \ref{guiding}a, scaled up by a factor of 100.
A marked overall signal-increase is observed in the presence of fields, mainly due to a transverse confinement of the molecules. 
In addition to the overall signal increase, the TOF distributions in fig.\ \ref{guiding}a are strongly structured because of the moving chain of traps. 
The many spikes in the experimental TOF trace, which are very well reproduced in the calculation, are formed because molecules are slowed down and bunched together as they move over the saddle points between traps.
Hence, the probability of finding a particle on a saddle point is higher, yielding higher intensity in the TOF trace.
The time difference between spikes is given by $\frac{\mathrm{L}}{2\mathrm{v}}=\frac{6 \mathrm{\, mm}}{\mathrm{v}}$, the time it takes the molecules to fly half a period of the potential. 
The TOF trace is dominated by one strong peak that corresponds to molecules flying at 288 m/s.
The width of this peak is again $\sim$$\frac{6 \mathrm{\, mm}}{\mathrm{v}}$, indicating that the principal trap is completely filled.
The specific shape of the peak of the guided molecules, shown enlarged in fig.\ \ref{guiding}b, is determined by the value of $\phi$ when the molecules are picked up: since the molecular packet is initially smaller than one trap, the pick-up efficiency changes when $\phi$ is varied between 0 and $\pi$, depending on whether the packet hits a trap or a saddle point. 
Experimental verification of this effect was achieved by changing the time delay $\Delta$t between the laser pulse and the start of the HV waveform. 
In fig.\ \ref{guiding}b molecules were guided at 288 m/s but $\phi$ was decreased in steps of $\sim$$\pi/4$ (bottom to top trace; positive traces are experimental data, negative traces are simulations): For a guiding frequency of 24 kHz, $\phi=\pi$ corresponds to $\Delta$t=20.83 $\mu$s. 
On the right of the TOF traces the initial phase-space distributions are sketched and overlaid on the separatrix for guiding (see below).
For $\Delta$t=0 and 20 $\mu$s, the first trap is exactly generated as the molecules enter, resulting in a high pickup efficiency. 
In contrast, at $\Delta$t=10 $\mu$s the molecules hit a saddle point and are spread over two traps.
The three local maxima in this peak indicate the times-of-flight corresponding to the edges of the two traps.

Figs.\ \ref{guiding}c and d show the simulated transverse and longitudinal acceptance of the device for the guiding of \CO using 16 kV p-p waveform amplitudes. 
Dots in these panels represent initial phase-space coordinates of molecules that were successfully guided through a structure with an artificial length of 20 m.
Lines show the separatrix, the edge of the volume outside of which no molecules are accepted.
The three separatrices in panel c are understood by inspecting the contour plots in fig.\ \ref{expt}b; radially, the trap changes between two configurations: a deep trap when it is located in the plane of an electrode, and a less deep trap when it is located exactly between two electrodes.
The innermost separatrix in fig.\ \ref{guiding}c corresponds to the latter case, the outermost one corresponds to the former case.
The dotted line shows an averaged case, calculated by integrating the radial potential along the structure.
Traps are transversely filled all the way out to the electrodes, and transverse velocities of $\pm$13 m/s are accepted. 
Most accepted molecules are inside the dashed separatrix, indicating that the simple integration is a good approximation.
Molecules outside this separatrix can still fly through the entire structure if their transverse oscillation is such that they happen never to find an exit.
The longitudinal separatrix shown in panel d confirms the pattern observed in the TOF traces: spatially, it fills the entire space between two traps.
At the deepest point, velocities $\pm$17 m/s are accepted. 
For \COn, the total calculated six-dimensional acceptance, as derived from trajectory simulations at 550 m/s, amounts to 0.3 10$^{-3}$ m$^3$ [m/s]$^3$.
For the guiding of OH(X $^2\Pi_{3/2}$, $v$=0, $J$=3/2, $M_J\Omega$=-9/4) the acceptance of the new decelerator is calculated, in the same way, to be 1.0 10$^{-3}$ m$^3$ [m/s]$^3$, compared to $\sim$0.35 10$^{-3}$ m$^3$ [m/s]$^3$ for the Stark decelerator \cite{Scharfenberg:2009p3004}.
For both structures, the acceptance is reduced when decelerating, as is shown in fig.\ \ref{guiding}e where the calculated acceptance for OH is plotted as a function of deceleration.
Here, trajectories are calculated for the deceleration from 550 m/s to 180 m/s with adjusted decelerator length to obtain different decelerations \cite{Scharfenberg:2009p3004}.
Without deceleration, the acceptance of the present decelerator is three times larger, and the absolute reduction at increasing deceleration is similar for both decelerators; the pseudopotential that has to be added to the longitudinal potential due to the acceleration -- and thus also the reduction of the barrier height in forward direction -- is the same in both cases.
Therefore, the present structure allows for either much more compact machines than the one from ref.\ \cite{Scharfenberg:2009p3004}, or for significantly increased acceptance at a given decelerator length.
Velocities below 180 m/s were avoided in the simulations in ref.\ \cite{Scharfenberg:2009p3004} due to the difficulties described above; since these difficulties are completely absent here, a comparison at low velocities would be even more favorable for the new decelerator.

\begin{figure}
 \includegraphics[width=3.2 in]{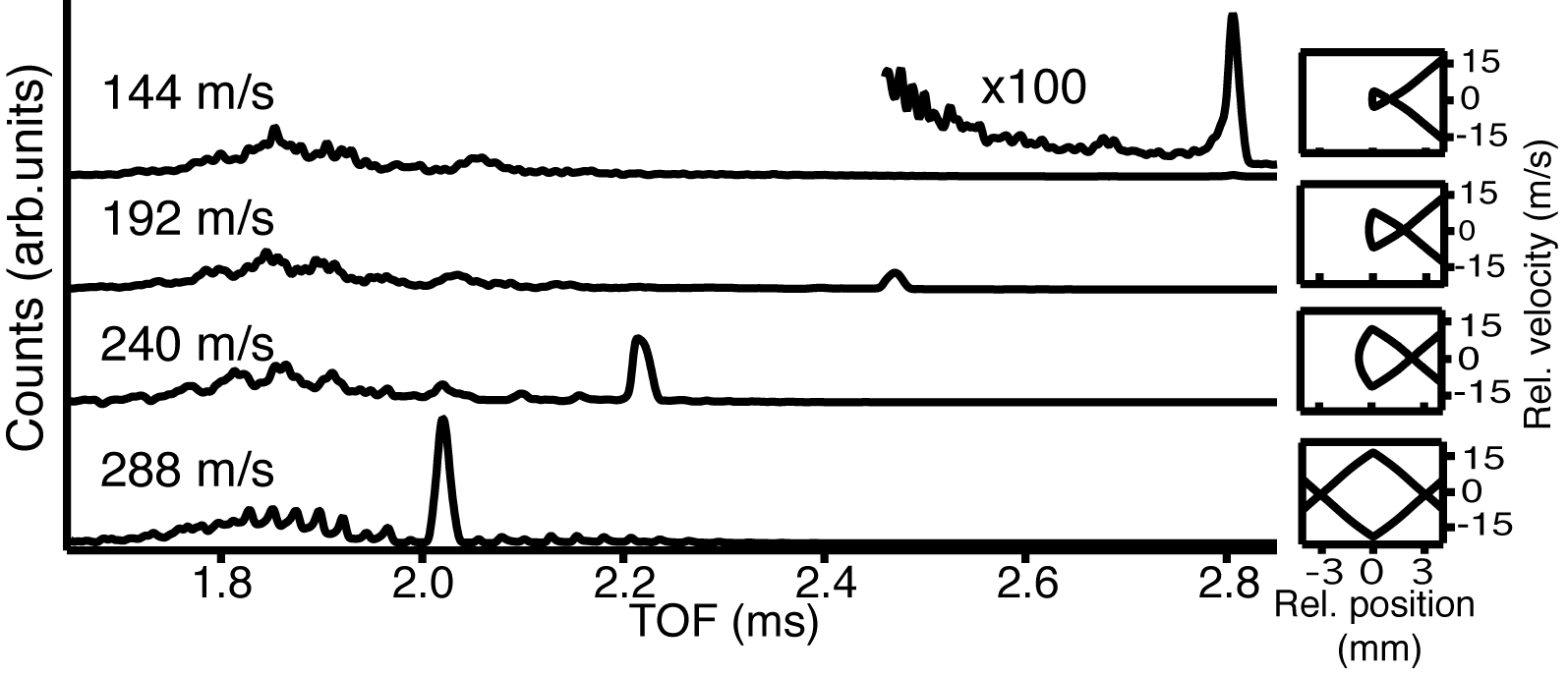}
\caption{\label{deceleration} Experimental TOF traces for decelerated CO. The initial velocity is 288 m/s in each case, final velocities are 144, 192, 240, and 288 m/s (from above). The insets on the right side of the figure show the longitudinal separatrices for the different cases of deceleration.}
 \end{figure}
Deceleration using chirped frequencies is demonstrated in fig.\ \ref{deceleration}. 
For all these measurements, the initial frequency was 24 kHz (288 m/s), and the amplitude was 16 kV p-p. 
While the molecules were inside the decelerator, the frequency was decreased to 20, 16, or 12 kHz (corresponding to 240, 192, and 144 m/s; the lowest attainable velocity is currently given by the electronics). 
The insets show the separatrices for each of the deceleration measurements.
Because of the reduced acceptance, the peak intensity is also reduced for higher accelerations.
Nevertheless, a deceleration from 288 m/s to 144 m/s could be observed with a good signal-to-noise ratio. 
For this measurement, an acceleration of $\sim$65 km/s$^2$ is exerted, and 75\% of the kinetic energy is removed.

In conclusion, a new kind of Stark decelerator has been demonstrated that employs continuously modulated electric fields to create a cylindrically symmetric traveling potential.
It has been shown that molecules can be guided at constant velocity, and they can also be decelerated. 
The large acceptance for a comparatively compact structure, make it an attractive alternative to a conventional Stark decelerator. 
A particularly attractive feature is the fact that the molecules are already trapped in three dimensions during deceleration. 
Consequently, when the modulation frequency is reduced to zero, molecules are confined in all spatial dimensions, thus rendering explicit trap loading obsolete. 
The new decelerator provides an ideal starting point for experiments like, for example, a molecular fountain, a storage ring (both for the loading of a ring, or even a ring-structure itself), high-resolution spectroscopy, or crossed beam collision studies.
An additional feature of the new decelerator is the fact that the molecules never move into regions where the electric field exceeds 30 kV/cm.
In heavy diatomic molecules, like for example YbF, the low rotational states that are low-field-seeking at low electric fields are turned into high-field-seeking states at strong fields due to their interaction with rotationally excited states \cite{Tarbutt:2009p4094}.
So far, such molecules could only be decelerated using alternating gradient techniques.
These methods are, however, not very efficient, and experimentally very challenging.
In contrast, the present approach will directly allow for the deceleration of these molecules in an environment where they never move in regions of strong electric field.

\begin{acknowledgments}
We greatly acknowledge the help provided by Georg Heyne and Viktor Platschkowski in the development and implementation of the electronics.\end{acknowledgments}


\end{document}